\documentclass{osa-article}

\journal{osajournal}

\articletype{Research Article}

\usepackage{braket}

\newcommand{\change}[1]{#1}

\begin{document}

\title{\change{Localization effects from local phase shifts in the modulation of waveguide arrays}}

\author{Konrad Tschernig,\authormark{1,*} 
        Armando Perez-Leija,\authormark{1} 
				and 
				Kurt Busch\authormark{2,3}}

\address{\authormark{1}CREOL, The College of Optics and Photonics, 
				               University of Central Florida, Orlando, FL 32816, USA\\
		\authormark{2}Max-Born-Institut, 
                       12489 Berlin, Germany\\
         \authormark{3}Humboldt-Universit\"{a}t zu Berlin, Institut f\"{u}r Physik, 
				               AG Theoretische Optik \& Photonik, 12489 Berlin, Germany}

\email{\authormark{*}konrad.tschernig@knights.ucf.edu} 



\begin{abstract}
Artificial gauge fields enable the intriguing possibility to manipulate the propagation 
of light as if it were under the influence of a magnetic field even though photons possess 
no intrinsic electric charge. Typically, such fields are engineered via periodic modulations 
of photonic lattices such that the effective coupling coefficients after one period become 
complex-valued. In this work, we investigate the possibility to introduce randomness into  
artificial gauge fields by applying local random phase shifts in the modulation of lattices 
of optical waveguides. We first study the elemental unit consisting of two coupled single-mode 
waveguides and determine the effective complex-valued coupling coefficient after one period 
of the modulation as a function of the phase shift\change{, the modulation amplitude and the modulation frequency}. Thereby
we identify the regime where varying the modulation phase yields sufficiently large changes of 
the effective coupling coefficient \change{to induce Anderson localization}. Using these results, we demonstrate numerically the onset 
of Anderson localization in \change{1D- and 2D-}lattices of $x$-, and helically-modulated waveguides via randomly 
choosing the modulation phases of the individual waveguides. 
Besides further fundamental investigations of wave propagation in the presence of random gauge fields,
our findings enable the engineering of the coupling coefficients without changing the footprint 
of the overall lattice. \change{As a proof of concept, we demonstrate how to engineer out-of-phase modulated lattices which exhibit dynamic localization and defect-free surface states.} Therefore, we anticipate that the modulation phase will play an 
important role in the judicious design of functional waveguide lattices.
\end{abstract}

\section{Introduction}

Anderson localization 
\cite{PhysRev.109.1492}
is the phenomenon of wave localization via multiple scattering in disordered systems. 
More precisely, for sufficiently strong disorder, interference processes change the 
physical nature of waves from propagating to localized so that Anderson localization 
represents a universal phenomenon and occurs in numerous wave systems
\cite{RevModPhys.80.1355,lagendijk2009fifty}, 
including optics, acoustics, and crystals. 
Historically, Anderson localization has first been studied in electronic systems where
additional effects such as electron-electron interaction and the application of 
magnetic fields modify the characteristics of the Anderson transition, thereby providing
a rich field for both fundamental research (e.g. field theories) and exciting applications 
(e.g. quantum-Hall-effect based devices) \cite{Ying2016}. 
In fact, Anderson localization has also been demonstrated in classical wave systems 
such as acoustics 
\cite{Macon1991,maynard2001acoustical}, optics 
\cite{Lahini2008,Martin2011,Giuseppe2013,segev2013anderson}, and in crytals \cite{Ying2016}.
For instance, arrays of coupled optical waveguides 
\cite{christodoulides2003discretizing},
ring resonators \cite{bogaerts2012silicon}
or optical cavities 
\cite{walter2016classical}
are described via tight-binding models where the the on-site (diagonal) 'potentials'
are given by the individual waveguides' propagation constants or resonance frequencies 
of the individual resonators/cavities. By the same token, the (off-diagonal) 'hopping' 
elements between sites are determined by the modal overlap of the waveguide modes or
the resonator/cavity modes.
For instance, in arrays of coupled waveguides both on- and off-diagonal disorder can 
be readily induced and controlled 
\cite{segev2013anderson}
by manipulating the effective refractive index of the individual waveguides \cite{Lahini2008} and the 
distance between them \cite{Martin2011}.\par
Nonetheless, the aforementioned richness of electronic systems appears to be out of 
reach for optical systems, since electron-electron interaction and the effects of magnetic 
fields generally require the presence of electric charges.
However, there exists an intriguing possibility to modify the flow of light via 
artificial gauge fields 
\cite{aidelsburger2018artificial}. 
Such fields can be induced via periodic modulation of photonic lattice structures, 
for example by changing the transverse positions of a waveguide array in a periodic 
fashion 
\cite{jorg2020artificial}. 
Then, periodic modulations in the waveguide positions induce complex-valued off-diagonal 
coupling parameters ('hopping' elements), which are directly associated to an artificial 
gauge field that acts on the propagation of light (see section \ref{sec:gauge-fields}). 
Notably, this effect occurs despite the fact that photons possess no intrinsic electric 
charge. 
Such artificial gauge fields have been exploited to implement, for example, photonic 
Floquet topological insulators 
\cite{rechtsman2013photonic} 
and to guide light in novel fashions on different optical platforms 
\cite{lin2014light,umucalilar2011artificial,schmidt2015optomechanical,lim2017electrically,tsesses2018optical,shen2022generation,westerberg2016synthetic,longhi2013effective}. \par

With a view on Anderson localization, the need to control the gauge field locally emerges 
and has been the subject of several recent studies in the context of resonator lattices 
\cite{fang2012realizing,schmidt2015optomechanical}. 
In this work, we study theoretically and numerically the possibility of generating locally 
random gauge fields in modulated waveguide arrays. We achieve this by introducing random 
phase shifts (modulation phases) along the propagation direction into the modulation of the 
waveguide positions in regular \change{1D- and 2D-}lattices. When the waveguides are modulated in a helical
fashion or along the lattice axis, we find that disorder in the modulation phases translates directly
into disorder in the effective coupling coefficients. Beyond \change{Anderson} localization our 
findings on the modulation phase open up a new pathway to engineer the coupling behavior 
in photonic lattices. Specifically, our approach allows for increasing and decreasing the 
coupling as desired without changing the average distance between the waveguides and, thus, 
without affecting the overall footprint of the photonic lattice. \change{As a proof of concept, we obtain and tune dynamic localization effects \cite{PhysRevB.34.3625,PhysRevLett.96.243901} by choosing the modulation phases judicously.}
\begin{figure*}[b]
\begin{center}
	\includegraphics[width=6in]{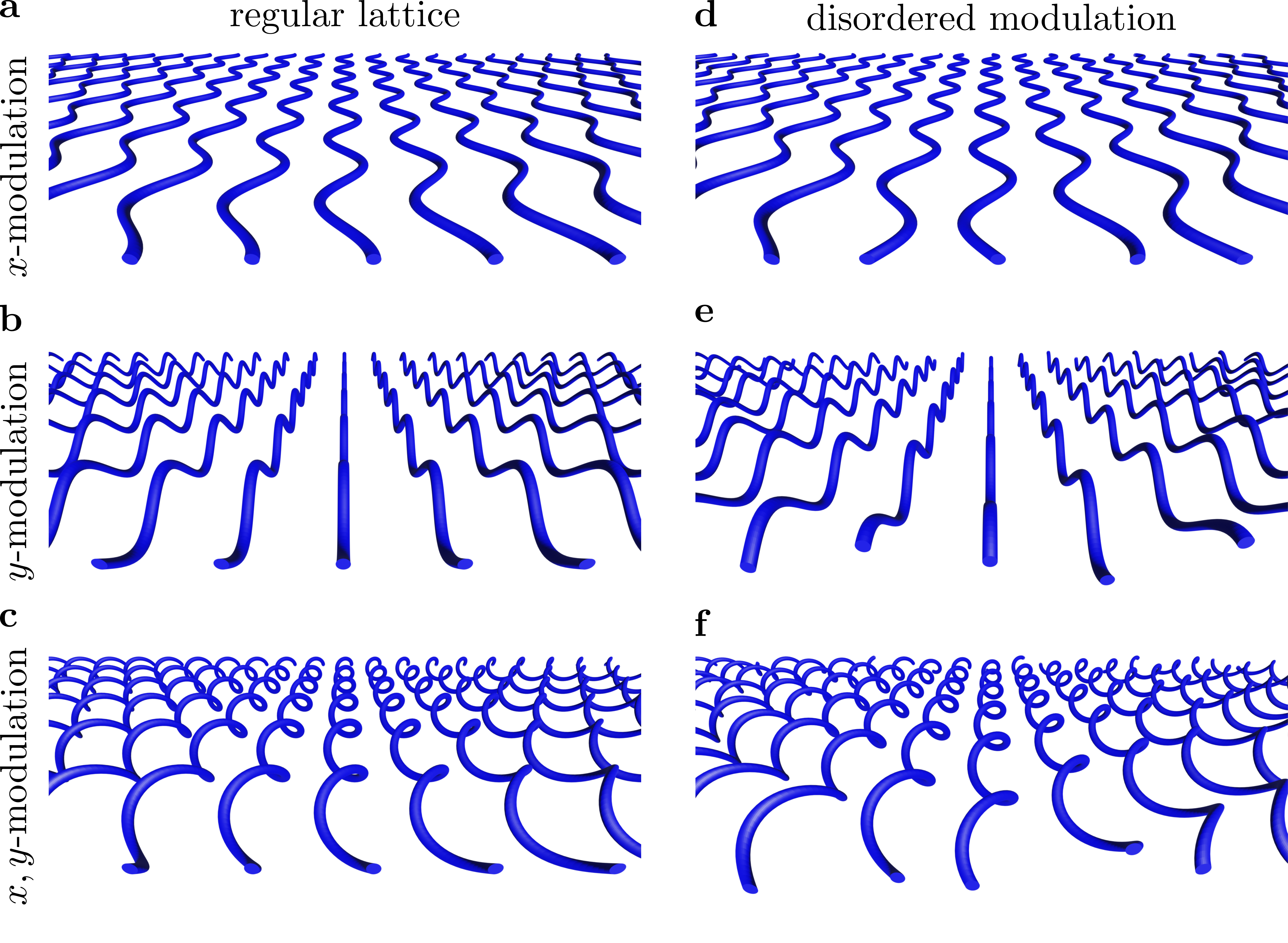}
\end{center}
\caption{\textbf{Modulated waveguide arrays exhibiting artifical gauge fields.} 
         \textbf{Panels a-c:} Regular lattices of waveguides subjected to $x$-, $y$- 
				                      and helical modulation, respectively. 
				 \textbf{Panels d-f:} Illustration of random phase-shifts in the modulation, 
				                      and thereby, disorder in the artificial gauge field.
			  }
\label{fig:lattices}
\end{figure*}

\section{Artificial gauge fields from periodic modulations}
\label{sec:gauge-fields}

Let us first elaborate on the origin of artificial gauge fields via periodic modulation of 
waveguide arrays. To set the stage, we begin with the Helmholtz equation in paraxial 
approximation 
\cite{kiselev2007localized}
for waves propagating in $z$-direction
\begin{equation}
	i\partial_z \psi(x,y,z) 
	= 
	\left[-\frac{1}{2k_0}\left(\partial_x^2+\partial_y^2 \right) +V \right]\psi(x,y,z),
	\label{eq:paraxial_eq}
\end{equation}
where $k_0=2\pi/\lambda$ is the wave number, $\psi(x,y,z)$ is the slowly-varying field 
envelope, $x,y$ the transverse coordinates, $z$ the propagation distance and the potential 
$V$ encodes the refractive index profile. We start by considering a lattice of $N$ straight 
waveguides, such that $V=V(x,y)=\sum_{n=1}^N V_0(x-x_n,y-y_n)$ and $V_0(x,y)$ is the refractive 
index profile of a single straight waveguide centered at the origin $(x,y)=(0,0)$, and 
$(x_n,y_n)$ is the transverse position of the $n$'th waveguide. 
To introduce periodic modulations, we shift the waveguide centers according to
$(x_n,y_n) \rightarrow(x_n+R_x \cos(\omega z),y_n+R_y \sin(\omega z))$, where $R_{x/y}$ 
are the modulation radii and $\omega$ is the modulation frequency along the propagation
direction. 
Evidently, this introduces a $z$-dependence into the potential 
$V(x,y)\rightarrow V(x,y,z)=\sum_{n=1}^N V_0(x-R_x\cos(\omega z)-x_n,y-R_y\sin(\omega z)-y_n)$. 
However, by transforming into the co-rotating reference frame 
$(x',y',z')=(x-R_x\cos(\omega z),y-R_y\sin(\omega z),z)$ we can eliminate the $z$-dependence 
in the potential and, instead, acquire an artificial gauge in the new coordinate system
\begin{equation}
	i\partial_{z'} \psi(x',y',z')
	=
	\left[-\frac{1}{2k_0}\left(\vec{\nabla}-i\vec{A}(z')\right)^2 +V(x',y') +\Phi(z')\right] \psi(x',y',z'),
	\label{eq:paraxial_corotating}
\end{equation}

where $\vec{\nabla}=(\partial_{x'},
\partial_{y'})^T$ and $\vec{A}(z')=k_0 \omega (R_x \sin(\omega z'), -R_y \cos(\omega z'))^T$ 
is the artificial gauge field. \change{Additionally, we acquire the $z'$-dependent term
$\Phi(z')=-\frac{1}{2}k_0\omega^2\big(R_x^2\sin^2(\omega z')$ $+R_y^2\cos^2(\omega z') \big)$, which commutes with all the other terms on the right hand side of Eq.~(\ref{eq:paraxial_corotating}), since it is independent of the transverse coordinates. Therefore, it will only contribute a global phase factor $\exp\left(-i\int_0^{z'} \Phi(z'') \mbox{d}z''\right)$  to the solution $\psi(x',y',z')$ and will not contribute to the output light intensities $|\psi(x',y',z')|^2$.} 
Crucially, the periodically modulated 
waveguide array is now equivalent to a lattice of straight waveguides in which the light 
is subjected to an artificial external magnetic field $\vec{B}=\vec{\nabla}\times\vec{A}$. 
To further elucidate the impact of the artificial gauge field on the evolution dynamics, we 
arrange the waveguides in a regular lattice, see Fig.~(\ref{fig:lattices}-a,b,c) along the 
$x$-axis, $(x_n,y_n)\rightarrow (n a,0)$, where $a$ is the lattice constant, and perform 
a Peierls' substitution 
\cite{PhysRev.84.814}
to obtain the tight-binding equation
\begin{equation}
i\partial_z \psi_m = \sum_{n=1}^N H_{m,n}(z) \psi_n.
\end{equation}
Here $\psi_m$ is the amplitude of the guided mode of the $m$'th waveguide and $H_{m,n}(z)$ 
denotes the $z$-dependent tight-binding Hamiltonian describing the hopping of the light-amplitudes 
among the lattice sites. \change{Note that we have dropped the ``prime'' on the coordinates for the sake of brevity.} In particular we assume the waveguides to be identical, such that 
$H_{n,n}=\beta=0$, and only allow nearest neighbor coupling $H_{n,n+1}(z)=\kappa e^{i\phi(z)}=H_{n+1,n}^*(z)$ 
($H_{n,m}(z)=0$ otherwise). Here, $\kappa$ is the real \change{and positive} coupling coefficient between nearest 
neighbors and $\phi(z)=\vec{A}(z)\cdot\vec{a}=a k_0 R_x \omega \sin(\omega z)$ is the effective 
artificial gauge field and $\vec{a}=(a,0)^T$ is the lattice vector. \change{In first-order approximation we can write the effective $z$-independent Hamiltonian $\hat{H}_{\text{eff}}$, which governs the evolution over one period $Z$, as $\hat{H}_{\text{eff}} = 1/Z \int_0^Z \hat{H}(z) \text{d}z$ \cite{Eckardt_2015}, which is the average over one preiod. Thus, we find the effective coupling coefficient
\begin{equation}
\kappa_{\text{eff}}=\frac{\kappa}{Z} \int_0^Z e^{i a k_0 R_x \omega \sin(\omega z)} \text{d}z = \kappa J_0(a k_0 R_x \omega),
\label{eq:bessel}
\end{equation}
which is a well-known result \cite{Eckardt_2015}, and implies that the effective coupling strength between the waveguides can be controlled by changing the nearest-neighbor distance, wave length, modulation radius and frequency. In what follows we introduce and study an additional degree of freedom in the form of a relative phase shift in the modulation of the waveguides.}

\section{Random artificial gauge fields}
\begin{figure*}[b]
\begin{center}
	\includegraphics[width=\linewidth]{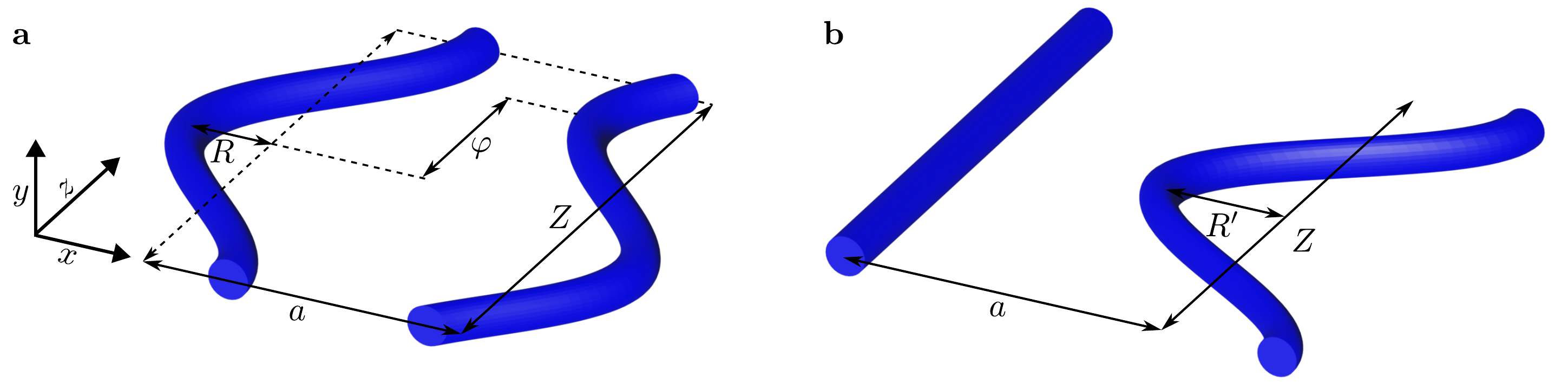}
\end{center}
\caption{\textbf{Two modulated coupled waveguides.}
         \textbf{Panel a:} In the laboratory frame of reference both waveguides 
				                   are modulated with period $Z$, radius $R$ and relative 
													 phase shift $\varphi$. 
													 Note, the center points of the modulation are separated 
													 by the distance $a$ (lattice constant). 
				 \textbf{Panel b:} When transformed into the rest-frame of the first waveguide 
				                   the system is equivalent to a single straight waveguide 
													 coupled to a modulated waveguide with the effective modulation 
													 radius $R'=2 R \sin(\varphi/2)$ subjected to an artificial 
													 gauge field.}
\label{fig:two_waveguides}
\end{figure*}
We now study the effects of adding local random phase shifts $\varphi_n$ to the periodic modulation 
of the waveguide positions, as we show in Fig.~(\ref{fig:lattices}-d,e,f). Therefore, we consider 
the $z$-dependent potential of the form 
$V(x,y,z)=\sum_{n=1}^N V_0(x-R_x\cos(\omega z+\varphi_n)-x_n,y-R_y\sin(\omega z+\varphi_n)-y_n)$. 
In other words, each waveguide performs the same transverse movement as before but is now displaced 
along the $z$-direction by the amount $z_n=\varphi_n/\omega$. Clearly, there is no single reference 
frame within which the potential is rendered completely $z$-independent. Then the question arises 
immediately, whether a gauge field can even exist under such circumstances. We answer this positively 
and to see that, we first consider a pair of waveguides $N=2$ as we show in Fig.~(\ref{fig:two_waveguides}). 
Without loss of generality we choose the first waveguide as the reference and set $\varphi_1=0$ and 
relabel $\varphi_2\rightarrow\varphi$. Following the recipe outlined above, we transform into the 
co-rotating frame of reference of the first waveguide 
$(x',y',z')=(x-R_x\cos(\omega z),y-R_y\sin(\omega z),z-\frac{\varphi}{2\omega})$ and arrive at the 
transformed potential $V(x',y',z')=V_0(x',y')+V_0(x'-a+R_x'\sin(\omega z'),y'-R_y'\cos(\omega z'))$, 
with the effective modulation radii $R_{x/y}'=2R_{x/y}\sin(\varphi/2)$ and observe that the $z'$-dependence 
of the second waveguide potential remains. To take the movement of the second waveguide into account, 
we model the real-valued coupling coefficient $\kappa$ as an exponentially decaying function of 
their distance $d=|\vec{a}|$, \change{$\kappa(z')=\kappa_0 e^{-\xi |\vec{a}(z')|} e^{i\phi(z')}$}, 
where $\kappa_0$ and $\xi$ are \change{real and positive} fit-parameters. Note that we obtain the same 
vector potential \change{as before except for a phase shift $\vec{A}(z',\varphi)=k_0\omega (R_x \sin(\omega z'+\varphi/2,-R_y\cos(\omega z' +\varphi/2))$. Additionally, the displacement vector $\vec{a}(z',\varphi)=(2R_{x}\sin(\varphi/2)\sin(\omega z')-a,2R_{y}\sin(\varphi/2)\cos(\omega z'))^T$ between the waveguides is now a function 
of $z'$ and $\varphi$. Thus, in first-order approximation, we expect from our theory that the effective coupling coefficient over one period to be a function of the modulation phase
\begin{equation}
\kappa_{\text{eff}}(\varphi)=\frac{\kappa_0}{Z} \int_0^Z e^{-\xi |\vec{a}(z',\varphi)|}e^{i\vec{A}(z',\varphi)\cdot \vec{a}(z',\varphi)}\mbox{d}z'.
\end{equation}
}
 
\par
Let us now verify these theoretical considerations. \change{Our goal is to obtain the effective (tight-binding) coupling coefficient after one period of the modulation and to study its dependence on the modulation phase. To this end, we employ the beam propagation 
method (BPM) to solve the paraxial equation\change{, Eq.~(\ref{eq:paraxial_eq})}, numerically. From the BPM simulations we obtain the evolution operator $\hat{U}(Z)$ and extract the effective coupling from it. First, we setup two waveguides at distance $a$ and modulate their positions as described above. Next, we launch the guided mode $\ket{\psi_n(0)}$ of each waveguide, $n=1,2$, to obtain the corresponding output distribution $\ket{\psi_n(Z)}$ after one period of the modulation.} Evidently, the input and output fields are related by 
$\ket{\psi_n(Z)}=\hat{U}(Z)\ket{\psi_n(0)}$, which enables us to find the matrix elements of 
the propagation operator via
\begin{figure*}[b]
\begin{center}
  \includegraphics[width=\linewidth]{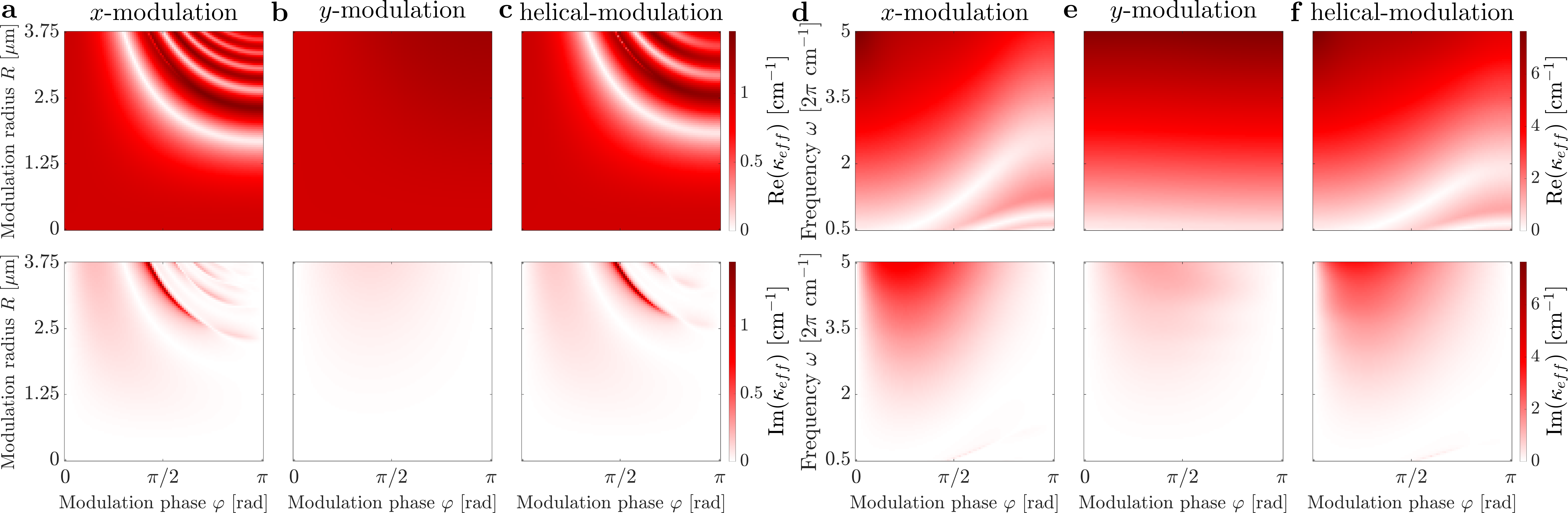}
\end{center}
\caption{\change{\textbf{Parameter scans over the modulation phase $\varphi$ versus modulation radius $R$ (\textbf{a-c}) and versus the modulation frequency $\omega$ (\textbf{d-f}).} 
            For each parameter pair we performed BPM simulations over one period $Z=2\pi/\omega$ (wavelength $\lambda=450$ nm, distance 
						between modulation centers $a=10$ $\mu$m, waveguide diameter $d=2.5$ $\mu$m, relative 
						refractive index contrast $\Delta n_0=1.55 \cdot 10^{-3}$) of the waveguide dimer 
						and extracted the effective coupling coefficient $\kappa_{\text{eff}}$, for 
						\textbf{a, d} $x$-modulation, 
						\textbf{b, e} $y$-modulation and 
						\textbf{c, f} helical modulation. 
						The top (bottom) panels show the real (imaginary) part of $\kappa_{\text{eff}}$. In the scans over the radii we have chosen $\omega=2\pi/cm$ and in the scans over the frequencies we chose $R=2.5 \mu$m.}}
\label{fig:scans}
\end{figure*}

\begin{equation}
	U_{m,n}(Z)\coloneqq \bra{\psi_m(0)}\hat{U}(Z)\ket{\psi_n(0)}=\braket{\psi_m(0)|\psi_n(Z)},
\end{equation}
where the inner product is defined as $\braket{\psi|\phi}=\sum_{p,q} \phi_{p,q} \psi_{p,q}^*$ 
and $\psi_{p,q}$ is the field amplitude on the (numerical) discretization point with indices $p,q$. 
\change{To find the effective coupling coefficient $\kappa_{\text{eff}}=\kappa e^{i\phi}$ we assume that the effective Hamiltonian acquires the form
\begin{equation}
\hat{H}_{\text{eff}}=\begin{pmatrix}
0 & \kappa e^{i\phi} \\
\kappa e^{-i\phi} & 0
\end{pmatrix}.
\label{eq:Heff}
\end{equation}
By identifying $\hat{U}(Z)=e^{iZ\hat{H}}$ and after some algebra we find $\kappa=\frac{\omega}{2\pi}\arctan\left(|\lambda_+ +\lambda_-|/|\lambda_+-\lambda_-| \right)$ and $\phi=\arccos(1-|\vec{v}_+^T \cdot \vec{v}_-|^2)/2$, where $\lambda_{\pm}$ ($\vec{v}_\pm$) are the eigenvalues (eigenvectors) of $\hat{U}(Z)$.
}

We now perform parameter scans over the modulation radius $R \in [0,R_{\text{max}}]$, \change{the modulation frequency $\omega \in [\pi/cm,10\pi/cm]$} and the
 modulation phase $\phi \in [0,\pi]$ and obtain the corresponding effective coupling coefficient 
$\kappa_{\text{eff}}$ for three different cases of modulation: $x$- ($R_x=R,R_y=0$), 
$y$- ($R_x=0, R_y=R$) and helical modulation ($R_x=R=R_y$). Note, that we choose 
$R_{\text{max}}=(a-d)/2$, where $d$ is the waveguide diameter, so that the waveguides do 
not overlap during any point of the propagation. \change{Furthermore, we avoid too small modulation periods to minimise bending losses.} Our results are summarized in Figs.~(\ref{fig:scans}).
First of all, we observe that in the case of $y$-modulation, Fig.~(\ref{fig:scans}-b,e), that 
the coupling between the waveguides is almost completely independent of the modulation radius 
and phase. Thus, we can expect that random disorder in the modulation phase will not be able to 
induce Anderson localization under $y$-modulation for any modulation radius $R$. On the other 
hand, for $x$- and helical modulation, we find a significant change of the coupling when 
$R\gtrsim 2 \mu$m over the whole range of the modulation phase \change{and frequency}. 

\section{Anderson localization from locally disordered gauge fields}

\begin{figure*}[b]
\begin{center}
	\includegraphics[width=\linewidth]{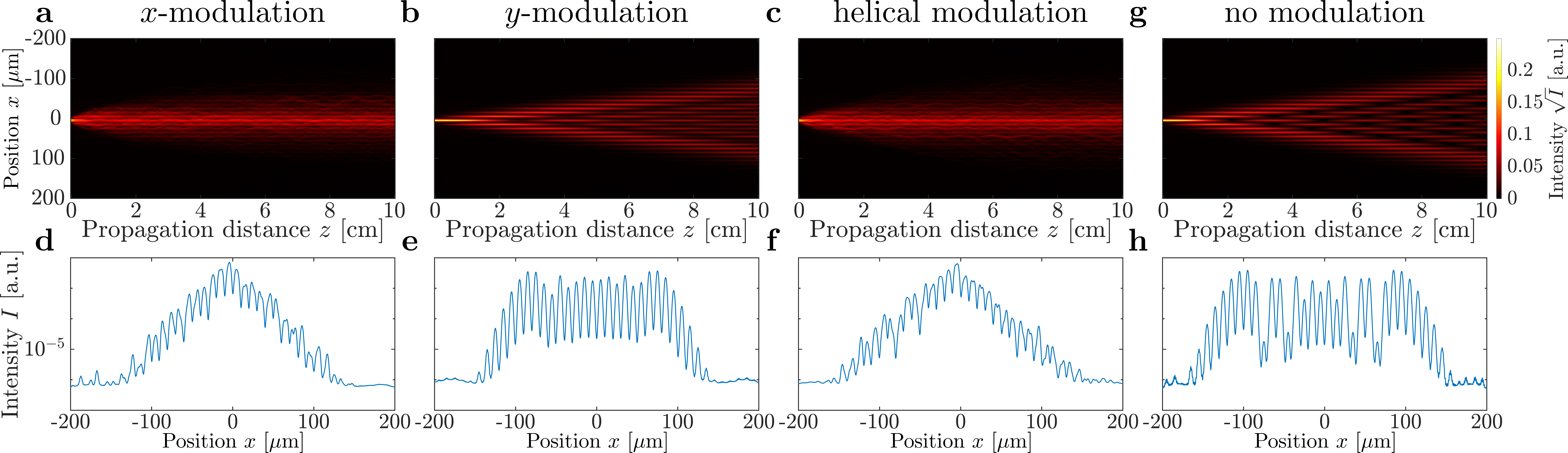}
\end{center}
\caption{\textbf{Anderson localization from random modulation phases.} 
         \textbf{a-c} Intensity evolution averaged over 50 instances of the randomly chosen 
				              modulation phases $\varphi_n \in [0,2\pi]$ for $x$-, $y$-, and helical 
											modulation respectively. In all cases we set the corresponding modulation 
											radius to $R=2.5 \mu$m. 
				 \textbf{d-f} Logarithm of the output intensity distributions. In the case of $x$- 
				              and helical modulation, we observe the signatures of Anderson localization 
											in the exponential decay of the averaged intensity along the transverse 
											direction $x$. For $y$-modulation, the disorder in the modulation phases 
											does not introduce sufficiently strong disorder in the effective coupling 
											coefficient to localize the light. 
				\change{\textbf{g-h} For comparison we show the evolution of a single-site excitation in a lattice of straight waveguides without any modulation or disorder.} }
\label{fig:anderson}
\end{figure*}

Now we are in the position to study Anderson localization due to disorder in the 
modulation phase. In order to observe Anderson localization, we require sufficiently strong 
disorder in the coupling coefficients. As we have shown in the previous section, we expect 
to achieve this using $x$- and helical modulation with a modulation radius $R=2.5 \mu$m \change{and a modulation frequency of $\omega=2\pi/cm.$ Let us first study the 1D-case of linear arrays of waveguides and then the case of 2D lattices.\par}
We now launch light into the center of a $N=40$ waveguide array and let it propagate for 
$z=10$ cm and take the average over an ensemble of $N_{\text{ens}}=50$ instances of the 
randomly chosen modulation phases $\varphi_n \in [0,2\pi]$. In Fig.~(\ref{fig:anderson}-a,b,c), 
we show the evolution of the ensemble-averaged intensity in the disordered lattices. 
Note, we show the integrated intensity $I(x,z)=\int  |\psi(x,y,z)|^2 \text{d}y$ and also 
chose to plot $\sqrt{I}$ in order improve the visibility of small intensities. We observe localization of the light in the $x-$ and helical modulation, while the light 
exhibits the usual discrete diffraction behaviour in the case of $y$-modulation\change{, compare with Fig.~(\ref{fig:anderson}-g,h) where we show the discrete diffraction pattern in a lattice of straight waveguides without disorder}. To further 
corroborate our findings we show the output intensity as a logarithmic plot in Fig.~(\ref{fig:anderson}-d,e,f). 
Here, we observe a triangular shape of the intensity distribution which is characteristic 
of Anderson localization and indicates the exponential decay of the intensity in the 
transverse direction from the initial excitation site, which is completely absent for
$y$-modulation. \par
\begin{figure*}[t]
\begin{center}
	\includegraphics[width=\linewidth]{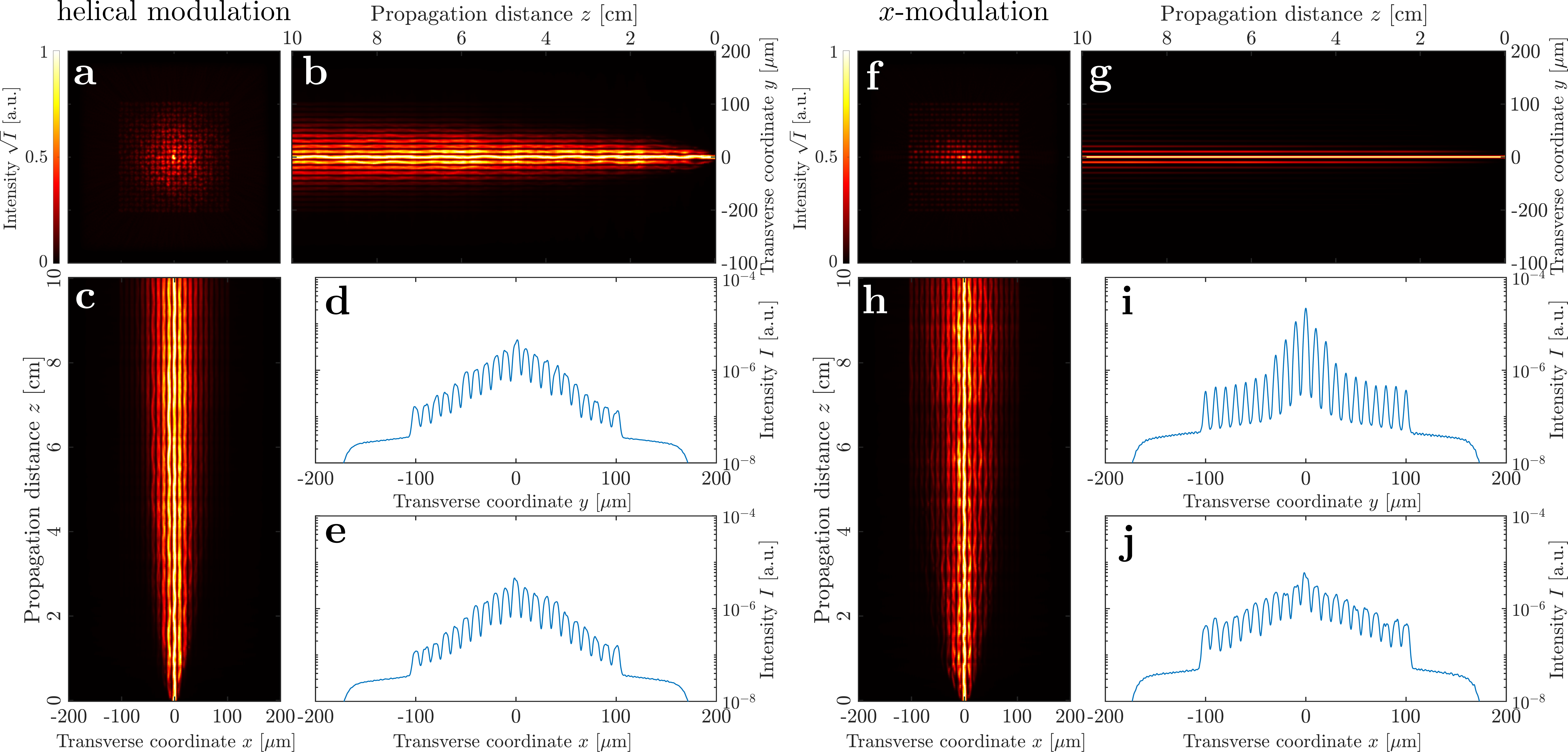}
\end{center}
\caption{\change{\textbf{2D Anderson localization from random modulation phases.} We show the ensemble-averaged intensity evolution in random-out-of-phase modulated 2D arrays of waveguides (square 21x21 lattice) under (\textbf{a-e}) helical modulation and (\textbf{f-j}) $x$-modulation. \textbf{a, f} Output intensity \textbf{b, g} Intensity along propagation direction (integrated over $x$-direction) \textbf{c, h} Intensity along propagation (integrated over $y$-direction) \textbf{d, e, i, j} Logarithmic plot of the output intensity along the $x$- ($y$-) direction. }}
\label{fig:anderson2d}
\end{figure*}
\change{In the light of these findings we now turn our attention to disordered 2D lattices. For our numerical experiments we set up a square lattice of $21 \times 21$ waveguides. We choose $a=10~\mu m$, $\omega=2\pi/cm$ and $R=2.5~\mu m$, facilitating sufficient disorder in the coupling when the modulation phases are chosen randomly. Due to the symmetry of this setup, we only need to consider the cases of helical- and $x$-modulation since $y$-modulation is now equivalent to $x$-modulation, apart from a trivial 90$^\circ$ rotation around the $z$-axis. We launch light in the center of the square lattice and obtain the average of the intensity evolution over $N_{\text{ens}}=20$ instances of the modulation disorder over a propagation distance of $z=10$ cm. In Fig.~(\ref{fig:anderson2d}) we depict the corresponding results. As can be seen in Fig.~(\ref{fig:anderson2d}-a-e), under helical modulation, the light experiences Anderson localization in both $x$- and $y$-direction. This is rather unsurprising, since in 2D the helical modulation is indistinguishable from $x$-modulation in the 1D case. The more interesting case is the $x$-modulation, as seen in Fig.~(\ref{fig:anderson2d}-f-j). From the previous 1D-results one might expect to see the simultaneous existence of Anderson localization (in $x$-direction) and discrete diffraction (in $y$-direction). While we indeed observe Anderson localization in the modulation direction, we encounter a completely different behaviour in the perpendicular direction. Firstly, we would like to note the logarithmic plot of the output intensities in Fig.~(\ref{fig:anderson2d}-i), which suggests an almost Gaussian distribution of the light rather than the exponentially localized ones in Fig.~(\ref{fig:anderson2d}-d,e,j). Further, it is worth to appreciate the degree of slowdown that the light experiences in its spreading in $y$-direction when compared to the $x$-direction. These observations raise the first question, of why we do not observe discrete diffraction in $y$-direction? To answer this, it is important to realize that the intensity distributions shown in Fig.~(\ref{fig:anderson2d}-g,i) are averaged not only over the ensemble but also over the $x$-coordinate. Thus, even before ensemble-averaging, we observe an incoherent superposition of diffraction processes that ``began'' at random positions along the $x$- and $z$-axis, thereby smoothing out the typical 1D-interference pattern shown in Fig.~(\ref{fig:anderson}-b,g). Furthermore, in this 2D setup we can expect that significant coupling between next-nearest neighbors in the diagonal directions occurs, thereby further inhibiting a clean diffraction pattern. While these explanations seem rather intuitive, they do not explain the significant narrowing of the intensity distribution and it is yet unclear how to properly characterize the process in $y$-direction. We conjecture that this behaviour constitutes either the early onset of Anderson localization (exhibiting a much larger localization length in $y$ than in $x$) or a slow diffusion-like process of light or maybe a mixture of the two. This question certainly merits further investigations. \par}

\change{
\section{Dynamic localization from out-of-phase modulation}
Until now we have utilized the modulation phase to introduce disorder in the local effective gauge fields and the resulting effective coupling coefficients in waveguide lattices. Another intruiging possibility is the judicious design of the local effective gauge field in order achieve a desired transformation. As a proof of concept we demonstrate how to obtain lattices exhibiting dynamic localization (DL), a well-known effect that occurs in periodically modulated lattices \cite{PhysRevB.34.3625,PhysRevLett.96.243901}. For example, for in-phase modulated arrays, DL is achieved at the roots of the $J_0$ Bessel-function in Eq.~(\ref{eq:bessel}) and this constitutes the most prevalent case that has been studied in the past \cite{Guzman-Silva:20,PhysRevLett.96.243901}. Notably, Huang et. al \cite{Huang:13} report a DL effect due to the $\pi$-out-of-phase refractive index modulation in sub-wavelength-scale light-guiding structures. Here we demonstrate numerically how to obtain DL in $\varphi$-out-of-phase modulated structures. \par
\begin{figure*}[t]
\begin{center}
	\includegraphics[width=\linewidth]{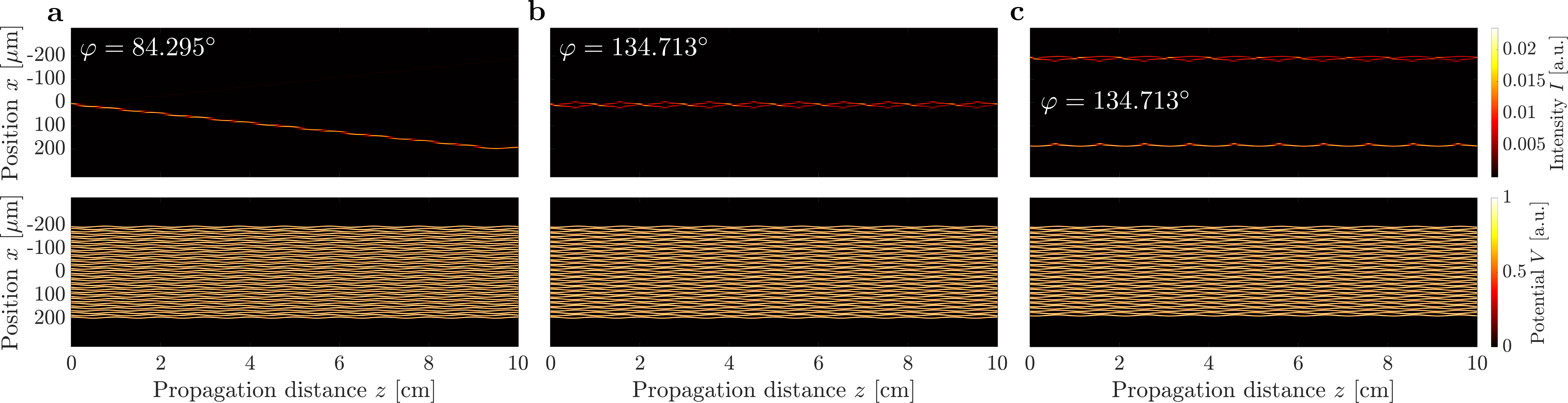}
\end{center}
\caption{\change{\textbf{Dynamic localization in out-of-phase $x$-modulated 1D waveguide arrays} We show the intensity evolution of dynamically localized single-site excitations. The modulation phases $\varphi_n$ are chosen to alternate between 0 and $\varphi$ ($a=10~\mu m$, $\omega=2\pi/cm$, $R=2.5~\mu m$). In order to achieve dynamic localization we choose $\varphi$ such that the effective coupling between adjacent waveguides is either (\textbf{a}) minimized, $\varphi=84.295^\circ$, obtaining ``shifting'' dynamic localization, or (\textbf{b}) maximized, $\varphi=134.713^\circ$, yielding standard dynamic localization. (\textbf{c}) With the lattice configured for standard dynamic localization we observe defect-free surface states. The top (bottom) panels show the evolution of the intensity (waveguide potential)}.}
\label{fig:dynamic_localization}
\end{figure*}
To obtain DL, the effective coupling coefficient after one period must vanish. This means that a single-site excitation launched in waveguide $n$ will return to the same waveguide after one period, although the wavefuntion may spread out to some extent in between. Thus we need to find the modulation parameters such that the effective coupling vanishes or is minimized. As an example we choose $x$-modulation, $R=2.5~\mu m$, $a=10~\mu m$ and $\omega=2\pi/cm$. Using Fig.~(\ref{fig:scans}-a) we find that the coupling is minimized for a modulation phase of $\varphi_d=84.295^\circ$. Now we need a 1D lattice of waveguides in which each nearest-neighbor pair of waveguides exhibits this relative modulation phase. This is achieved by choosing $\varphi_{2n}=0$ and $\varphi_{2n+1}=\varphi_d$, in other words, by alternating the modulation phase along the waveguide array. In Fig.~(\ref{fig:dynamic_localization}-a) we show the resulting lattice structure and the evolution of a single-site excitation of the central waveguide $n$. Quite surprisingly, we observe that, after one period, the light does not re-emerge at the initial waveguide $n$ but at waveguide $n+2$, thus traveling across the lattice during propagation. However, it can also be seen clearly that the wavepacket evolves without any significant diffraction. To the best of our knowledge this constitutes a new form of dynamic localization, which one may call ``shifting dynamic localization''. It is worth noting that it is possible to control the direction of the shifting localization by choosing either an even or odd injection site. This intriguing behavior can be understood in the following way. Firstly, the interaction between each pair of waveguides after one period can be regarded as a generalized waveguide beamsplitter as seen in Eq.~(\ref{eq:Heff}). Therefore, the condition of vanishing coupling implies that light entering either waveguide will emerge in the same waveguide at the output. In other words, in this configuration, the effective waveguide beamsplitter is fully transparent. However, the situation is slightly different in a lattice configuration. In our specific case the light couples completely into the opposite waveguide $n+1$ at roughly half of the modulation period. Coincidentally, at this point of the propagation, the next-nearest-neighbor waveguide is closest and thus the light continues to couple into waveguide $n+2$ instead of returning to waveguide $n$. This consideration leads to the natural question, as to what happens when the generalized beam splitter is fully reflecting rather than transparent? To answer this, we need to find the modulation phase that maximizes the effective coupling. Using Fig.~(\ref{fig:scans}-a) we find the maximum at $\varphi_d=134.713^\circ$. In Fig.~(\ref{fig:dynamic_localization}-b) we show the resulting lattice structure and the evolution of the central excitation. Now we clearly observe what is conventionally understood as dynamic localization, since the light returns to its injection site after each period of the modulation. It is worth to appreciate that the waveguide structures shown in Fig.~(\ref{fig:dynamic_localization}-a,b) are hardly distinguishable to the human eye, and yet they feature fundamentally distinct light coupling behaviour. \par
Another phenomenon, which is closely related to dynamic localization, is the existence of so-called defect-free surface states \cite{PhysRevLett.100.203904,PhysRevLett.101.203902} in modulated lattices. To be precise, when the lattice is configured to exhibit dynamic localization, then light injected into the edge of such a lattice will remain localized on the edge. As we show in Fig.~(\ref{fig:dynamic_localization}-c) we observe exactly this effect in our out-of-phase modulated arrays. The lattice shown here, is identical to the one in Fig.~(\ref{fig:dynamic_localization}-b) except we have removed the right-most waveguide, in order to show that the defect-free surface mode is observable with both types of lattice terminations.
}

\section{Summary \& Outlook} 
To summarize, we have shown that the modulation phase readily enables us to manipulate 
the local artificial gauge field in waveguide lattices. By adjusting the relative modulation 
phase between adjacent waveguides we are then able to tune the real- and imaginary part 
of the effective coupling coefficients. As a proof of concept we have demonstrated 
numerically the possibility to induce Anderson localization by choosing the modulation 
phases randomly \change{in both 1D- and 2D-lattices}. Since such waveguide lattices are straightforward to implement, our studies enable the 
testing of our fundamental understanding of wave propagation in disordered systems 
and to simulate other physical systems such as electron propagation in \change{random} magnetic fields. 
\change{Further, we have shown how to obtain dynamic localization effects by tuning the modulation phases in 1D-lattices. Notably, we found ``shifting'' dynamic localization, which enables the diffraction-free transport of single-site excitations across 1D lattices.} \par 

\change{Looking forward, our proposed technique of out-of-phase modulation may} be useful in the design of functional optical devices. 
As an example, the so-called $\hat{J}_x$-lattice, which performs the quite useful discrete 
fractional fourier-transform \cite{Weimann2016,Tschernig:18}, 
requires the coupling coefficients to follow a parabolic distribution. In this specific 
case, the conventional method, of engineering the distances between waveguides in 
order to achieve the desired coupled structure, leads to an unfavorable exponential growth 
in the distances between the waveguides. Our method, however, does not require any change 
in the lattice vector and a wide range of coupling profiles \change{may be achievable} by a judicious 
choice of the modulation phases. In this way, our method \change{may} help to reduce the footprint 
of optical coupled structures. \change{We believe this to be a promising direction for future studies.\par}
\change{Another interesting avenue for further investigation, would be the possibility to leverage the modulation phase in the design of topological insulator lattices. Artificial gauge fields already play a crucial role in this area of research, most prominently in the case of Floquet topological insulators. For example, the topological Haldane lattice has been implemented via the helical in-phase-modulation of a honeycomb lattice of optical wavegudies \cite{rechtsman2013photonic}, which induces a transversally-homogeneous artificial gauge field. Instead, with the modulation phase as a new degree of freedom, is it possible to obtain topological lattices exhibiting transversally-structured artificial gauge fields? We hope that our work will form a stepping stone for research in this exciting direction.\par}
Finally, in this work we have limited the modulation radius 
to a maximum value in order to avoid waveguide overlap during the modulation. It would be
 an intriguing possibility to relax this condition and to allow for partial waveguide 
overlap\change{, complete merging or even crossing} of the waveguides. Evidently, then the tight-binding 
approximation would break down and it is not entirely clear whether an effective Hamiltonian 
can be properly defined under these conditions. We plan to investigate all these possibilities 
in future work.

\section{Backmatter}

\begin{backmatter}
\bmsection{Funding}
\noindent
Deutsche Forschungsgemeinschaft (Project number 255652081)

\bmsection{Acknowledgments}
\noindent
The authors acknowledge support by Deutsche Forschungsgemeinschaft (DFG) within the priority program
SPP 1839 Tailored Disorder (BU 1107/12-2, PE 2602/2-2). The authors thank Miguel A. Bandres for tremendously insightful discussions and suggestions.

\bmsection{Disclosures}
\noindent
The authors declare no conflicts of interest.

\bmsection{Data Availability Statement}
\noindent
Data underlying the results presented in this paper are not publicly available at this time 
but may be obtained from the authors upon reasonable request.


\end{backmatter}

\bibliography{literature}

\begin{thebibliography}{10}
\newcommand{\enquote}[1]{``#1''}

\bibitem{PhysRev.109.1492}
P.~W. Anderson, \enquote{Absence of diffusion in certain random lattices,}
  {\protect\JournalTitle{Phys. Rev.}} \textbf{109}, 1492--1505 (1958).

\bibitem{RevModPhys.80.1355}
F.~Evers and A.~D. Mirlin, \enquote{Anderson transitions,}
  {\protect\JournalTitle{Rev. Mod. Phys.}} \textbf{80}, 1355--1417 (2008).

\bibitem{lagendijk2009fifty}
A.~Lagendijk, B.~Van~Tiggelen, and D.~S. Wiersma, \enquote{Fifty years of
  anderson localization,} {\protect\JournalTitle{Phys. today}} \textbf{62},
  24--29 (2009).

\bibitem{Ying2016}
T.~Ying, Y.~Gu, X.~Chen, X.~Wang, S.~Jin, L.~Zhao, W.~Zhang, and X.~Chen,
  \enquote{Anderson localization of electrons in single crystals:
  {L}i$_{x}${F}e$_{7}${S}e$_{8}$,} {\protect\JournalTitle{Science Advances}}
  \textbf{2}, e1501283 (2016).

\bibitem{Macon1991}
L.~Macon, J.~P. Desideri, and D.~Sornette, \enquote{Localization of surface
  acoustic waves in a one-dimensional quasicrystal,}
  {\protect\JournalTitle{Phys. Rev. B}} \textbf{44}, 6755--6772 (1991).

\bibitem{maynard2001acoustical}
J.~D. Maynard, \enquote{Acoustical analogs of condensed-matter problems,}
  {\protect\JournalTitle{Reviews of modern physics}} \textbf{73}, 401 (2001).

\bibitem{Lahini2008}
Y.~Lahini, A.~Avidan, F.~Pozzi, M.~Sorel, R.~Morandotti, D.~N. Christodoulides,
  and Y.~Silberberg, \enquote{Anderson localization and nonlinearity in
  one-dimensional disordered photonic lattices,} {\protect\JournalTitle{Phys.
  Rev. Lett.}} \textbf{100}, 013906 (2008).

\bibitem{Martin2011}
L.~Martin, G.~D. Giuseppe, A.~Perez-Leija, R.~Keil, F.~Dreisow, M.~Heinrich,
  S.~Nolte, A.~Szameit, A.~F. Abouraddy, D.~N. Christodoulides, and B.~E.~A.
  Saleh, \enquote{Anderson localization in optical waveguide arrays with
  off-diagonal coupling disorder,} {\protect\JournalTitle{Opt. Express}}
  \textbf{19}, 13636--13646 (2011).

\bibitem{Giuseppe2013}
G.~Di~Giuseppe, L.~Martin, A.~Perez-Leija, R.~Keil, F.~Dreisow, S.~Nolte,
  A.~Szameit, A.~F. Abouraddy, D.~N. Christodoulides, and B.~E.~A. Saleh,
  \enquote{Einstein-podolsky-rosen spatial entanglement in ordered and anderson
  photonic lattices,} {\protect\JournalTitle{Phys. Rev. Lett.}} \textbf{110},
  150503 (2013).

\bibitem{segev2013anderson}
M.~Segev, Y.~Silberberg, and D.~N. Christodoulides, \enquote{Anderson
  localization of light,} {\protect\JournalTitle{Nature Photonics}} \textbf{7},
  197--204 (2013).

\bibitem{christodoulides2003discretizing}
D.~N. Christodoulides, F.~Lederer, and Y.~Silberberg, \enquote{Discretizing
  light behaviour in linear and nonlinear waveguide lattices,}
  {\protect\JournalTitle{Nature}} \textbf{424}, 817--823 (2003).

\bibitem{bogaerts2012silicon}
W.~Bogaerts, P.~De~Heyn, T.~Van~Vaerenbergh, K.~De~Vos, S.~Kumar~Selvaraja,
  T.~Claes, P.~Dumon, P.~Bienstman, D.~Van~Thourhout, and R.~Baets,
  \enquote{Silicon microring resonators,} {\protect\JournalTitle{Laser \&
  Photonics Reviews}} \textbf{6}, 47--73 (2012).

\bibitem{walter2016classical}
S.~Walter and F.~Marquardt, \enquote{Classical dynamical gauge fields in
  optomechanics,} {\protect\JournalTitle{New Journal of Physics}} \textbf{18},
  113029 (2016).

\bibitem{aidelsburger2018artificial}
M.~Aidelsburger, S.~Nascimbene, and N.~Goldman, \enquote{Artificial gauge
  fields in materials and engineered systems,} {\protect\JournalTitle{Comptes
  Rendus Physique}} \textbf{19}, 394--432 (2018).

\bibitem{jorg2020artificial}
C.~J{\"o}rg, G.~Queralt{\'o}, M.~Kremer, G.~Pelegr{\'\i}, J.~Schulz,
  A.~Szameit, G.~von Freymann, J.~Mompart, and V.~Ahufinger,
  \enquote{Artificial gauge field switching using orbital angular momentum
  modes in optical waveguides,} {\protect\JournalTitle{Light: Science \&
  Applications}} \textbf{9}, 1--7 (2020).

\bibitem{rechtsman2013photonic}
M.~C. Rechtsman, J.~M. Zeuner, Y.~Plotnik, Y.~Lumer, D.~Podolsky, F.~Dreisow,
  S.~Nolte, M.~Segev, and A.~Szameit, \enquote{Photonic floquet topological
  insulators,} {\protect\JournalTitle{Nature}} \textbf{496}, 196--200 (2013).

\bibitem{lin2014light}
Q.~Lin and S.~Fan, \enquote{Light guiding by effective gauge field for
  photons,} {\protect\JournalTitle{Physical Review X}} \textbf{4}, 031031
  (2014).

\bibitem{umucalilar2011artificial}
R.~Umucal{\i}lar and I.~Carusotto, \enquote{Artificial gauge field for photons
  in coupled cavity arrays,} {\protect\JournalTitle{Physical Review A}}
  \textbf{84}, 043804 (2011).

\bibitem{schmidt2015optomechanical}
M.~Schmidt, S.~Kessler, V.~Peano, O.~Painter, and F.~Marquardt,
  \enquote{Optomechanical creation of magnetic fields for photons on a
  lattice,} {\protect\JournalTitle{Optica}} \textbf{2}, 635--641 (2015).

\bibitem{lim2017electrically}
H.-T. Lim, E.~Togan, M.~Kroner, J.~Miguel-Sanchez, and A.~Imamo{\u{g}}lu,
  \enquote{Electrically tunable artificial gauge potential for polaritons,}
  {\protect\JournalTitle{Nature communications}} \textbf{8}, 1--6 (2017).

\bibitem{tsesses2018optical}
S.~Tsesses, E.~Ostrovsky, K.~Cohen, B.~Gjonaj, N.~Lindner, and G.~Bartal,
  \enquote{Optical skyrmion lattice in evanescent electromagnetic fields,}
  {\protect\JournalTitle{Science}} \textbf{361}, 993--996 (2018).

\bibitem{shen2022generation}
Y.~Shen, E.~C. Mart{\'\i}nez, and C.~Rosales-Guzm{\'a}n, \enquote{Generation of
  optical skyrmions with tunable topological textures,}
  {\protect\JournalTitle{ACS Photonics}} \textbf{9}, 296--303 (2022).

\bibitem{westerberg2016synthetic}
N.~Westerberg, C.~Maitland, D.~Faccio, K.~Wilson, P.~{\"O}hberg, and E.~M.
  Wright, \enquote{Synthetic magnetism for photon fluids,}
  {\protect\JournalTitle{Physical Review A}} \textbf{94}, 023805 (2016).

\bibitem{longhi2013effective}
S.~Longhi, \enquote{Effective magnetic fields for photons in waveguide and
  coupled resonator lattices,} {\protect\JournalTitle{Optics letters}}
  \textbf{38}, 3570--3573 (2013).

\bibitem{fang2012realizing}
K.~Fang, Z.~Yu, and S.~Fan, \enquote{Realizing effective magnetic field for
  photons by controlling the phase of dynamic modulation,}
  {\protect\JournalTitle{Nature photonics}} \textbf{6}, 782--787 (2012).

\bibitem{PhysRevB.34.3625}
D.~H. Dunlap and V.~M. Kenkre, \enquote{Dynamic localization of a charged
  particle moving under the influence of an electric field,}
  {\protect\JournalTitle{Phys. Rev. B}} \textbf{34}, 3625--3633 (1986).

\bibitem{PhysRevLett.96.243901}
S.~Longhi, M.~Marangoni, M.~Lobino, R.~Ramponi, P.~Laporta, E.~Cianci, and
  V.~Foglietti, \enquote{Observation of dynamic localization in periodically
  curved waveguide arrays,} {\protect\JournalTitle{Phys. Rev. Lett.}}
  \textbf{96}, 243901 (2006).

\bibitem{kiselev2007localized}
A.~Kiselev, \enquote{Localized light waves: Paraxial and exact solutions of the
  wave equation (a review),} {\protect\JournalTitle{Optics and Spectroscopy}}
  \textbf{102}, 603--622 (2007).

\bibitem{PhysRev.84.814}
J.~M. Luttinger, \enquote{The effect of a magnetic field on electrons in a
  periodic potential,} {\protect\JournalTitle{Phys. Rev.}} \textbf{84},
  814--817 (1951).

\bibitem{Eckardt_2015}
A.~Eckardt and E.~Anisimovas, \enquote{High-frequency approximation for
  periodically driven quantum systems from a floquet-space perspective,}
  {\protect\JournalTitle{New Journal of Physics}} \textbf{17}, 093039 (2015).

\bibitem{Guzman-Silva:20}
D.~Guzman-Silva, M.~Heinrich, T.~Biesenthal, Y.~V. Kartashov, and A.~Szameit,
  \enquote{Experimental study of the interplay between dynamic localization and
  anderson localization,} {\protect\JournalTitle{Opt. Lett.}} \textbf{45},
  415--418 (2020).

\bibitem{Huang:13}
C.~Huang, X.~Shi, F.~Ye, Y.~V. Kartashov, X.~Chen, and L.~Torner,
  \enquote{Tunneling inhibition for subwavelength light,}
  {\protect\JournalTitle{Opt. Lett.}} \textbf{38}, 2846--2849 (2013).

\bibitem{PhysRevLett.100.203904}
I.~L. Garanovich, A.~A. Sukhorukov, and Y.~S. Kivshar, \enquote{Defect-free
  surface states in modulated photonic lattices,} {\protect\JournalTitle{Phys.
  Rev. Lett.}} \textbf{100}, 203904 (2008).

\bibitem{PhysRevLett.101.203902}
A.~Szameit, I.~L. Garanovich, M.~Heinrich, A.~A. Sukhorukov, F.~Dreisow,
  T.~Pertsch, S.~Nolte, A.~T\"unnermann, and Y.~S. Kivshar,
  \enquote{Observation of defect-free surface modes in optical waveguide
  arrays,} {\protect\JournalTitle{Phys. Rev. Lett.}} \textbf{101}, 203902
  (2008).

\bibitem{Weimann2016}
S.~Weimann, A.~Perez-Leija, M.~Lebugle, R.~Keil, M.~Tichy, M.~Gr{\"a}fe,
  R.~Heilmann, S.~Nolte, H.~Moya-Cessa, G.~Weihs, D.~N. Christodoulides, and
  A.~Szameit, \enquote{Implementation of quantum and classical discrete
  fractional fourier transforms,} {\protect\JournalTitle{Nature
  Communications}} \textbf{7}, 11027 (2016).

\bibitem{Tschernig:18}
K.~Tschernig, R.~de~J.~Le\'{o}n-Montiel, O.~S.~M. {n}a Loaiza, A.~Szameit,
  K.~Busch, and A.~Perez-Leija, \enquote{Multiphoton discrete fractional
  fourier dynamics in waveguide beam splitters,} {\protect\JournalTitle{J. Opt.
  Soc. Am. B}} \textbf{35}, 1985--1989 (2018).

\end{thebibliography}

\end{document}